\documentclass{aa}  

\usepackage{graphicx}
\usepackage{txfonts}
\usepackage{longtable}    
\newcommand{\hi}{{\rm H{\textsc{i}}\,}}

\newcommand{\hei}{{\rm He{\textsc{i}}\,}}
\newcommand{\heii}{{\rm He{\textsc{ii}}\,}}

\newcommand{\nitii}{{\rm N{\textsc{ii}}\,}}

\newcommand{\neiii}{{\rm Ne{\textsc{iii}}\,}}

\newcommand{\nevi}{{\rm Ne{\textsc{vi}}\,}}

\newcommand{\mgv}{{\rm Mg{\textsc{v}}\,}}

\newcommand{\mgvii}{{\rm Mg{\textsc{vii}}\,}}

\newcommand{\sulii}{{\rm S{\textsc{ii}}\,}}
\newcommand{\suliii}{{\rm S{\textsc{iii}}\,}}

\newcommand{\fevii}{{\rm Fe{\textsc{vii}}\,}}

\newcommand{\fex}{{\rm Fe{\textsc{x}}\,}}

\newcommand{\fexiii}{{\rm Fe{\textsc{xiii}}\,}}

\newcommand{\fexix}{{\rm Fe{\textsc{xix}}\,}}

\begin{document}

\title{Variability of the adiabatic parameter in monoatomic thermal and non-thermal plasmas\thanks{Tables 1 and 2 are only available in 
electronic form at the CDS via anonymous ftp to cdsarc.u-strasbg.fr (130.79.128.5) or via 
http://cdsweb.u-strasbg.fr/cgi-bin/qcat?J/A+A/}}


  \author{ Miguel A. de Avillez\inst{1,2} \and Gerv\'asio J.  Anela\inst{1} \and Dieter Breitschwerdt\inst{2}}
  
\institute{Department of Mathematics, University of \'Evora, R. Rom\~ao Ramalho 59, 7000 \'Evora, Portugal\\
	\email{mavillez@galaxy.lca.uevora.pt}
	\and
	Zentrum f\"ur Astronomie und Astrophysik, Technische Universit\"at Berlin, Hardenbergstrasse 36, D-10623 Berlin, 
	Germany
}

\date{Received Month Day, 2018; accepted Month Day, Year}

\titlerunning{The internal energy and the variability of the adiabatic parameter}
\authorrunning{de Avillez et al.}

  \abstract
   {Numerical models of the evolution of interstellar and integalactic plasmas often assume that the adiabatic parameter $\gamma$ (the 
   ratio of the specific heats) is constant (5/3 in monoatomic plasmas). However, $\gamma$ is determined by the total internal energy of 
   the plasma, which depends on the ionic and excitation state of the plasma. Hence, the adiabatic parameter may not be constant 
   across the range of temperatures available in the interstellar medium.}
   {We aim to carry out detailed simulations of the thermal evolution of plasmas with Maxwell-Boltzmann and non-thermal ($\kappa$ and 
   $n$) electron distributions in order to determine the temperature variability of the total internal energy and of the 
   adiabatic parameter. }
   {The plasma, composed of H, He, C, N, O, Ne, Mg, Si, S, and Fe atoms and ions, evolves under collisional ionization equilibrium 
   conditions, from an initial temperature of $10^9$ K. The calculations include electron impact ionization, radiative and dielectronic 
   recombinations and line excitation. The ionization structure was calculated solving a system of 112 linear equations using the Gauss 
   elimination method with scaled partial pivoting. Numerical integrations used in the calculation of ionization and excitation rates 
   are  carried out using the double-exponential over a semi-finite interval method. In both methods a precision of $10^{-15}$ is 
   adopted.}
   {The total internal energy of the plasma is mainly dominated by the ionization energy for temperatures lower than $8\times 10^4$ 
   K with the excitation energy having a contribution of less than one percent. In thermal and non-thermal plasmas composed of H, 
   He, and metals, the adiabatic parameter evolution is determined by the H and He ionizations leading to a profile in general having 
   three transitions. However, for $\kappa$ distributed plasmas these three transitions are not observed for $\kappa<15$ and for 
   $\kappa<5$ there are no transitions. In general, $\gamma$ varies from 1.01 to 5/3. Lookup tables of the $\gamma$ parameter are 
   presented as supplementary material.}
   {}

   \keywords{atomic processes -- atomic data -- hydrodynamics -- methods: numerical -- ISM general -- (galaxies): intergalactic 
   medium}

   \maketitle
%

\section{Introduction}

Numerical models of the interstellar and galactic media assume that (i) all the gas parcels in the plasma have the same ionic and 
radiative histories, (iii) the electrons in the plasma have a Maxwell-Boltzmann distribution (hereafter denoted by MB), and (iv) the 
adiabatic parameter $\gamma$ is a constant.

The first assumption implies that the ionic and radiative histories of the plasma are locked into the cooling function $\Lambda(T)$, 
which in turn is used as a sink term in the energy equation. $\Lambda(T)$ comprises the loss of energy through radiation by a gas parcel 
cooling from an initial temperature of $10^{8}-10^{9}$ K where it is assumed to be completely ionized and evolves under specific 
conditions: collisional ionization equilibrium (CIE) or non-equilibrium ionization (NEI) \citep[see discussions in, for 
example,][]{shapiro1976,schmutzler1993,sutherland1993,gnat2007,avillez2010}. 

In the interstellar and intergalactic gas simulations at each timestep the temperature of the gas is calculated over the computational 
domain. For each temperature a value of the cooling function is interpolated from lookup tables of $\Lambda(T)$ (erg cm$^{3}$ 
s$^{-1}$) normalized to $n_{H}n_{e}$ (hydrogen and electron number densities in cm$^{-3}$) previously calculated for an optically thin 
plasma. Therefore, there is no determination of the ionization structure of the plasma on the fly.

There is mounting evidence that frequently in low density plasmas electrons may be described by non-thermal distributions, for 
example, $\kappa$ \citep{vasyliunas1968}, $n$ \citep{hares1979, seely1987}, depleted high energy tail \citep{druyvesteyn1930, 
behringer1994}, and hybrid MB/power-law tail \citep{berezhko1999, porquet2001, dzifcakova2011}. These electron 
distributions occur in any place where a high temperature or density gradient exists, or when energy is deposited into the tail of the 
distribution at a rate that is sufficiently high to overcome the establishment of thermal equilibrium described by the 
MB distribution. For applicability in the Astrophysical context see discussions and references therein in 
\citet{wannawichian2003,karlicky2012,dzifcakova2013,nicholls2013,dudik2014,humphrey2014,avillez2015,avillez2017}.

The constancy of the adiabatic parameter used in the simulations is at odds with the ionic evolution of the plasma as its value 
depends on the plasma internal energy, which includes the contributions from the thermal, ionization and excitation energies \citep[see 
discussions in, e.g.,][]{cox1968, schmutzler1993}. Thus, a $\gamma$ parameter consistent with the underlying ionic and radiative 
histories must be included in these simulations \citep[for different applications of molecular and monoatomic gases see, e.g., 
][]{decampli1978,wuchterl1990,bodenheimer2013,vandenbroucke2013,vaidya2015}. Here we advance previous works by 
considering the evolution of the adiabatic parameter in plasmas characterized by thermal and non-thermal distributions ($\kappa$ and 
$n$) and provide lookup tables of the data that can be used in plasma simulations. 

The structure of this paper is as follows: Section 2 presents the non-thermal distributions used in the present work. Followed by a 
discussion of the internal energy of a gas parcel in Section 3. Section 4 deals with the framework associated to the thermal model 
adopted in the present calculations. In Section 5 results of the simulations are presented, while Section 6 describes 
the tabulated data, Section 7 closes the paper with some final remarks.

\section{Non-thermal distributions}

The $n-$ and $\kappa$ distributions in the energy space have the analytical forms \citep[][]{hares1979,seely1987}
\begin{equation}
f_{n}(E)dE=\frac{2}{\sqrt{\pi}(k_{B}T)^{3/2}}B_{n} E^{1/2}\left(\frac{E}{k_{B}T}\right)^{(n-1)/2}e^{-E/k_{B}T}dE,
\end{equation}
with
\begin{equation}
 \displaystyle B_{n}=\frac{\sqrt{\pi}}{2\Gamma(n/2+1)}\mbox{~and~} n\in[1,+\infty[.
 \end{equation}
and \citep{livadiotis2009,pierrard2010}
\begin{equation}
 f_{\kappa}(E)dE=\frac{2E^{1/2}}{\pi^{1/2}(k_{B}T)^{3/2}} \displaystyle A_{\kappa} \left[ 
 1+\frac{E}{(\kappa-3/2)k_{B}T}\right]^{-\kappa-1}dE,
\end{equation}
with 
\begin{equation}
A_{\kappa}=\frac{\Gamma(\kappa+1)}{\Gamma(\kappa-1/2)(\kappa-3/2)^{3/2}} \mbox{~and~} \kappa\in]3/2,+\infty[,
\end{equation}
respectively. In these expressions $E$ is the electron energy (eV), $k_{B}$ is the Boltzmann constant (erg K$^{-1}$), $T$ is the 
temperature (K), and $\Gamma(x)$ is the Gamma function of variable $x$. 

\begin{figure}[thbp]
	\centering
	\includegraphics[width=0.9\hsize,angle=0]{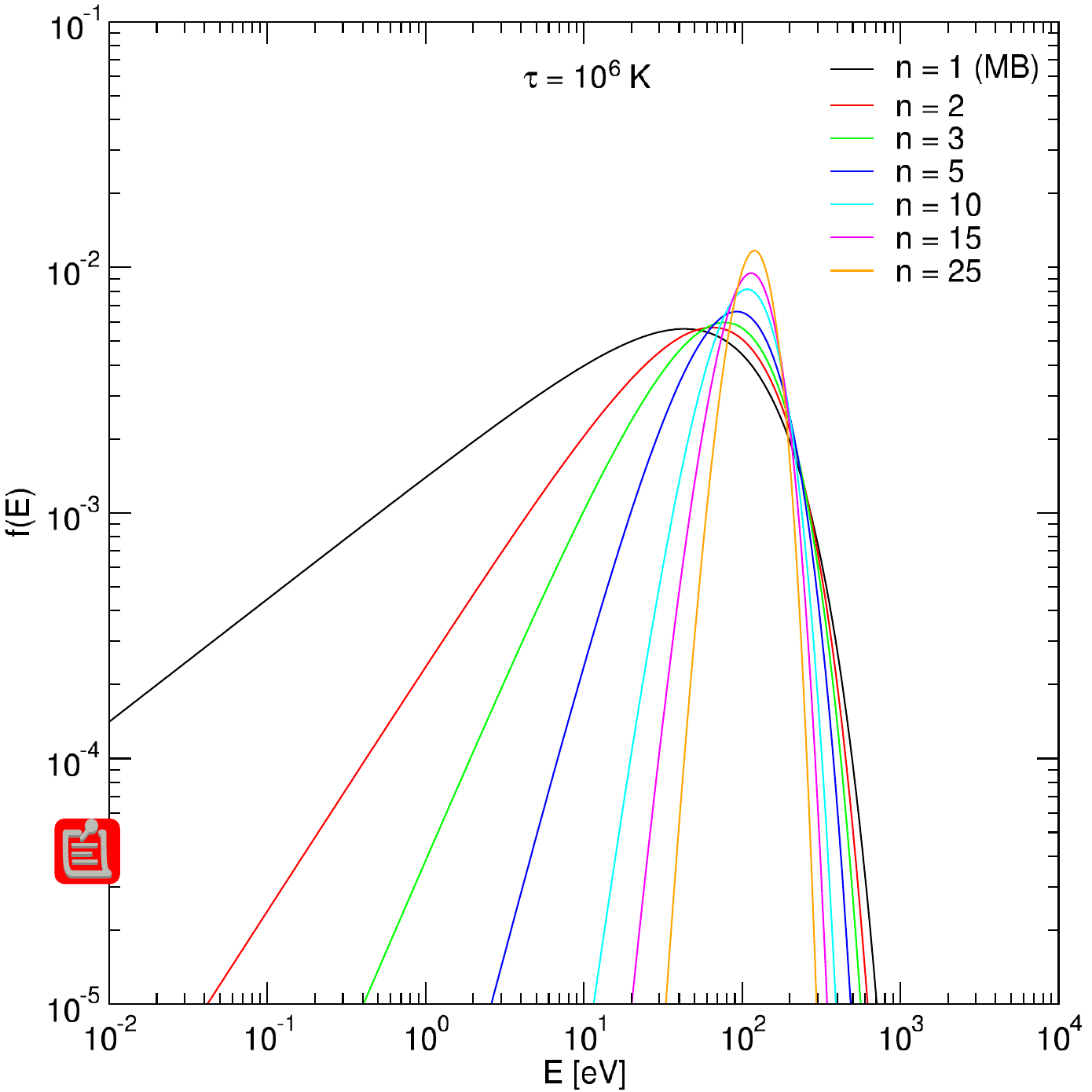}
	\includegraphics[width=0.9\hsize,angle=0]{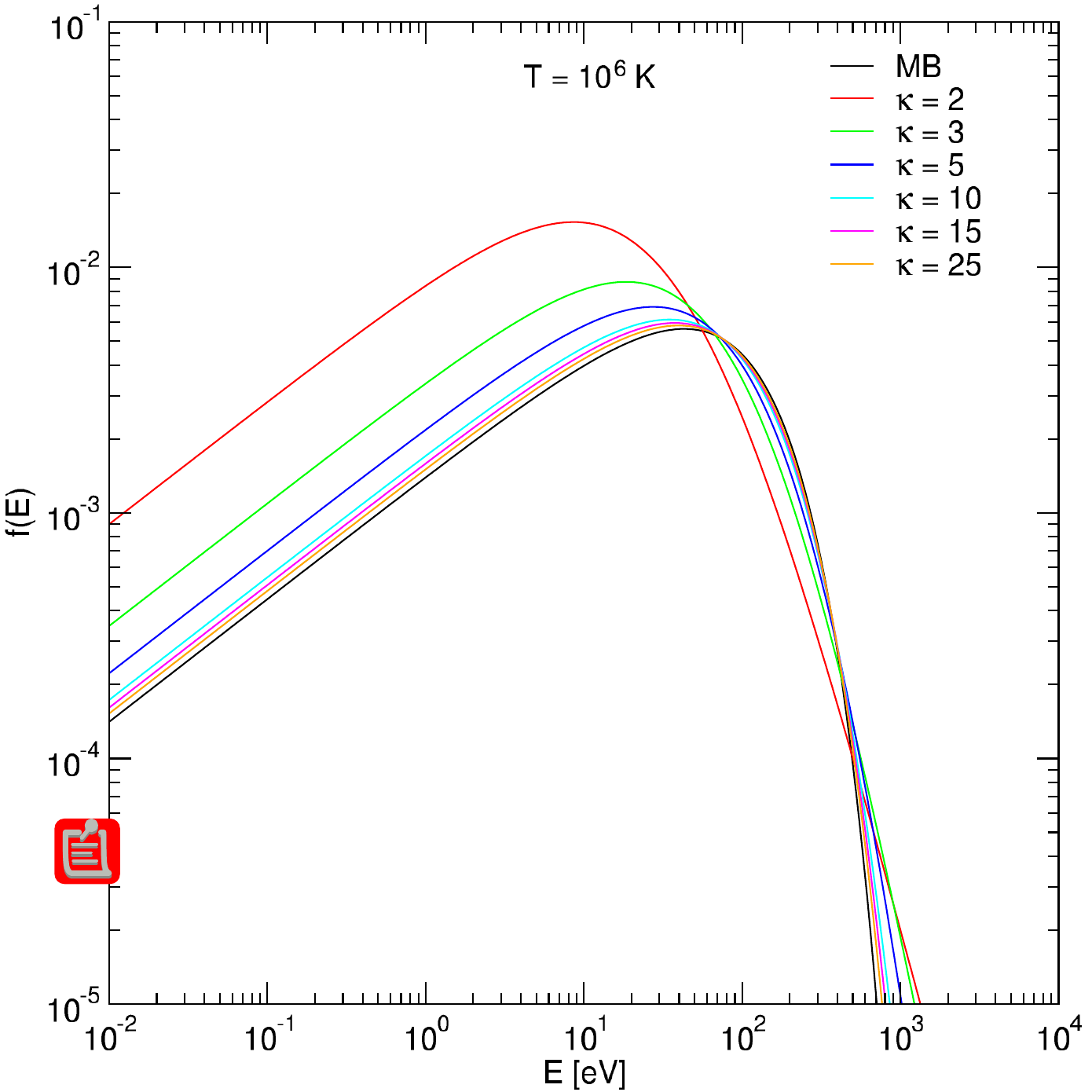}
	\caption{Maxwell-Boltzmann (both panels), $n$ (top panel) and $\kappa$ (bottom panel) distributions at $T=10^{6}$ K. The 
	$n>1$ distributions have a steeper high-energy tail than the MB distribution. The $\kappa$ distribution approaches MB when 
	$\kappa\to\infty$.}
	\label{distributions}
\end{figure}

The $n-$distribution (which becomes the MB distribution when $n=1$) is characterized by having a mean energy that depends 
on the $n$ parameter being given by $\langle E \rangle =(3/2)k_{B}\tau$ where $\tau=T(n+2)/3$ is a pseudo-temperature. This 
means that at $\tau$ the mean energies of the $n-$ and MB distributions are the same. Thus, $\tau$ has the same physical 
meaning as $T$ in the MB distribution. The $n-$distributions with the same $\tau$ have their peaks higher and narrower 
than that of the MB distribution, that is, they have less electrons with both high and low energies, but have an increased number of 
electrons with intermediate energies (top panel of Fig.~\ref{distributions}).

The $\kappa$-distribution is characterized by a high-energy power-law tail and having a mean energy that does not depend on 
$\kappa$ and is given by $\langle E\rangle=3/2k_{B}T$. Hence, $T$ can be defined as the thermodynamic temperature for these 
distributions. When $\kappa\to \infty,$ the Maxwell-Boltzmann distribution is recovered. As $\kappa$ decreases, deviations from 
the MB distribution increase, reaching a maximum when $\kappa$ approaches 3/2 (bottom panel of Fig.~\ref{distributions}). 

\section{Internal energy and the adiabatic parameter $\gamma$}

The internal energy of the gas parcel includes the contributions due to the thermal translational energy plus the energy stored in 
(or delivered) from the high ionization stages and from the excitation levels \citep[see,  
e.g.,][]{cox1968,macfarlane1989,schmutzler1993}
\begin{equation}
\label{energy}
\rho u=\frac{3}{2}(n_{tot}+n_{e}) k_{B}T+E_{si}+E_{se}\mbox{~~erg cm$^{-3}$}
\end{equation}
with $\rho$, $u$, $k_{B}$ and $T$ denoting the gas density (g cm$^{-3}$), the specific energy of the gas (erg g$^{-1}$), respectively; 
$n_{tot}$ and $n_{e}$ are the total number density of the element with atomic number $Z$ (hereafter referred as element $Z$) and the 
electron density, respectively, and are given by
\begin{equation}
n_{tot}=\sum_{Z}n_{_{Z}}=\sum_{Z}\sum_{z=0}^{Z}n_{Z,z} \mbox{~and~} n_{e}=\sum_{Z}\sum_{z=1}^{Z}z n_{_{Z,z}},
\end{equation}
where $z$ is the ion state, and $n_{_{Z}}$ and $n_{_{Z,z}}$ are the number densities of the element $Z$ and of its ionic state $z$. 
The energies stored in ionization and in excitation are respectively
\begin{equation}
E_{si}=\sum_{Z}\sum_{z=1}^{Z}n_{_{Z,z}}\left(\sum_{l=0}^{z-1}\Phi_{Z,l}\right) \mbox{~~erg cm$^{-3}$}
\end{equation}
and
\begin{equation}
\label{se}
E_{se}=\sum_{Z}\sum_{z=1}^{Z}\left[\sum_{j=n_{_{o}}+1}n_{_{Z,z,j}}\left(\Delta E_{j,n_{_{o}}}\right)_{_{Z,z}}\right]\mbox{~~erg 
cm$^{-3}$}.
\end{equation}
In these equations $\Phi_{Z,l}$ is the ionization potential (erg) of the ionization state $l$ of the element $Z$ and $\Delta E_{j,n_{_{o}}}$ is 
the excitation energy (erg) with respect to the ground level, $n_{_{o}}$, of each ion. If the gas is to be ionized and excited then $\rho u$ 
erg cm$^{-3}$ must be added to the gas. Of this amount $E_{si}$ strips the atoms and ions of their electrons, $E_{se}$ promotes the 
excitation of electrons in the atoms/ions and the reminder brings the system to a common temperature T. 

The plasma pressure is 
\begin{equation}
\label{pressure}
p=(n_{tot}+n_{e}) k_{B}T=n_{tot} \left(1+\langle Z\rangle\right) k_{B}T=\frac{\rho}{\mu m_{u}} \left(1+\langle Z\rangle\right)k_{B}T
\end{equation}
where $\langle Z\rangle=n_{e}/n_{tot}$ and $\mu$ are the mean electron charge and mean molecular weight, respectively. From 
(\ref{energy}) and (\ref{pressure}) it is seen that
\begin{equation}
p=\frac{2}{3}\rho u -E_{si}-E_{se}
\end{equation}
which is different than the classical relation $p=(\gamma-1)\rho u$ \citep[see, e.g.,][]{vandenbroucke2013}. Consequently, the  
adiabatic parameter $\gamma=(\partial \ln p/\partial \ln \rho)_{s}$ depends on the state of the plasma through the  internal energy. 
It can be written as \citep{cox1968}
\begin{equation}
\gamma=\frac{1}{c_{_{V}}}\left(\frac{p}{T \rho}\right)\chi_{T}^{2}+\chi_{\rho},
\end{equation}
with $c_{\textsc{v}}=\left(\partial u/\partial T \right)_{\textsc{v}}$ (erg g$^{-1}$ K$^{-1}$) denoting the specific heat at 
constant volume, and the coefficients $\chi_{T}$ and $\chi_{\rho}$ (temperature and density exponents) are written in terms of 
$\mu$ as
\begin{equation}
\chi_{T}=\frac{T}{p}\left(\frac{\partial p}{\partial T}\right)_{\rho}=1-\frac{T}{\mu}\left(\frac{\partial \mu}{\partial T}\right)_{\rho}
\end{equation}
and
\begin{equation}
\chi_{\rho}=\frac{\rho}{p}\left(\frac{\partial p}{\partial \rho}\right)_{T}=1-\frac{\rho}{\mu}\left(\frac{\partial \mu}{\partial 
	\rho}\right)_{T}.
\end{equation}

\section{Thermal model}
In order to study the evolution of $\gamma$ with temperature we follow the evolution of a gas parcel freely cooling from $10^{9}$ 
K (where it is assumed to be completely ionized)  evolving in collisional ionization equilibrium (CIE). The gas parcel is composed of the
ten most abundant elements in Nature (H, He, C, N, O, Ne, Mg, Si, S, and Fe) having solar abundances. These are based on 
\citet{asplund2009} and include the updates to the Mg, Si, S, and Fe abundances by \citet[][see also \citet{amarsi2017} for the 
abundance of Si]{scott2015a,scott2015b}. The physical processes included in these calculations comprise electron impact 
ionization (including excitation-autoionization), radiative and dielectronic recombinations and line excitation by electrons assuming 
the coronal approximation. In these calculations line excitation/de-excitation by protons, continuum emission and charge 
exchange reactions are not included.

\subsection{Ionic fractions}

The time-independent evolution of the ion fractions due to the 102 ions and 10 atoms (in a total of 112 equations) where  ionization 
and recombinations of ions of nuclear charge $Z$ occur between neighbouring ionization stages $z-1$, $z$ and $z+1$, is given by
\begin{equation}
\label{ionfracseq}
{\cal I}_{Z,z-1}x_{Z,z-1}n_{e}-({\cal I}_{Z,z}+ {\cal R}_{Z,z})x_{Z,z}n_{e}+{\cal R}_{Z,z+1}x_{Z,z+1}n_{e}=0,
\end{equation}
where ${\cal R}_{Z,z}$ and ${\cal I}_{Z,z}$ are the rates of recombination and ionization from state $(Z,z)$ to $(Z,z-1)$ and 
$(Z,z+1)$, respectively; $n_{e} $is the electron density and $x_{Z,z}=n_{Z,z}/n_{Z}$  the ionic fraction of element $Z$ with effective 
charge $z$. The ion density is then given by
\begin{equation}
n_{Z,z}=x_{Z,z}\,n_{Z}=x_{Z,z}\,A(Z)\,n_{H}
\end{equation}
with $A(Z)=n_{Z}/n_{H}$ and $n_{H}$ being the abundance of element $Z$ and the hydrogen density. The system of equations can 
be cast into the matrix equation
\begin{equation}
\mbox{AX}=0
\end{equation}
where $\mbox{X}$ is a vector comprising all ion fractions $x_{Z,z}$ and $\mbox{A}$ is a tridiagonal matrix with elements ${\cal         
I}_{Z,z-1}n_{e}$, $-({\cal I}_{Z,z}+ {\cal R}_{Z,z})n_{e}$, and ${\cal R}_{Z,z+1}n_{e}$ at each row populating the diagonal band.

\subsection{Ionization and recombination rates}

From the electron impact ionization cross sections we calculate the corresponding ionization rates associated to an ion of atomic 
number $Z$ and ionic charge $z$ by averaging the product $\sigma(E) v$ over the impacting particle kinetic energy distribution 
$f(E)$
\begin{equation}
\label{ionization_rate}
\langle \sigma v\rangle =\int_{I_{_{Z,z}}}^{+\infty} \sigma(E) (2E/m_{e})^{1/2} f(E) dE\mbox{~~cm$^{3}$ s$^{-1}$}
\end{equation}
where $m_{e}$ is the electron mass, and $I_{_{Z,z}}$ is the threshold energy in eV. 

The radiative and dielectronic recombination rates for the $n$ distribution are calculated from the fit coefficients to the Maxwellian 
rates \citep{dzifcakova1998}, that is, the radiative recombination rates for the $n$-distribution are determined from
\begin{equation}
\alpha_{RR}^{n}=\alpha_{_{Z,z}}^{MB}B_{n}\frac{\Gamma(n/2-\eta+1)}{\Gamma(3/2-\eta)}
\end{equation}
where $\alpha_{RR}^{MB}$ is the Maxwellian radiative recombination rate, and $\eta$ is a parameter of the power-law fit of the 
Maxwellian rate \citep[][]{woods1981}
\begin{equation}
\label{power_law}
\alpha_{RR}^{MB}=A_{rad}\left(\frac{T}{10^{4}\mbox{K}}\right)^{-\eta}\mbox{~~cm$^{3}$ s$^{-1}$}.
\end{equation}
The dielectronic recombination rates for the $n$-distribution of electrons are determined from coefficients of the Maxwellian rates 
given by the \citet{burgess1965} general formula
\begin{equation}
\alpha_{DR}^{MB}=\frac{1}{(k_{B}T)^{3/2}}\sum_{j}c_{j} e^{-E_{j}/(k_{B}T)} \mbox{~~cm$^{3}$ s$^{-1}$}
\end{equation}
through \citep{dzifcakova1998}
\begin{equation}
\label{dr_rate}
\alpha_{DR}^{n}=\frac{B_{n}}{(k_{B}T)^{3/2}}\sum_{j}c_{j} 
\left(\frac{E_{j}}{k_{B}T}\right)^{(n-1)/2}e^{-E_{j}/(k_{B}T)}\mbox{~~cm$^{3}$ s$^{-1}$.}
\end{equation}
In these expressions $A_{rad}$ and $c_{j}$ are fitting coefficients.

The radiative and dielectronic recombination rates for the $\kappa$ distribution are also based on the Maxwellian rates and are 
expressed respectively by \citep[][see also \citet{dzifcakova1992}]{wannawichian2003,dzifcakova2013}
\begin{equation}
\alpha_{RR}^{\kappa}=\alpha_{_{Z,z}}^{MB}A_{\kappa} \frac{\Gamma(\kappa+\eta-1/2)}{\Gamma(\kappa+1)}(\kappa-3/2)^{-\eta+3/2} 
\mbox{~~cm$^{3}$ s$^{-1}$} 
\end{equation}
and
\begin{equation}
\alpha_{DR}^{\kappa}=\frac{A_{\kappa}}{(k_{B}T)^{3/2}}\sum_{j}c_{j}\left[1+\frac{E_{j}}{(\kappa-3/2)k_{B}T}\right]^{-(\kappa+1)} 
\mbox{~~cm$^{3}$ s$^{-1}$} 
\end{equation}

\subsection{Levels populations and excitation rates}

The levels' population are calculated assuming excitation-deexcitation equilibrium, that is, equilibrium between the excitations (by 
particle impact) and deexcitations (by particle impact and spontaneous decays) to and from a level. We assume that the excitations and 
deexcitations by particle impact are only due to electrons. The population of level $j$ is determined from the collisional excitations of 
electrons from levels $m$ to level $j$ ($m<j$; excitation rate $C^{e}_{mj}$) and from level $j$ to levels $n$ ($j<n$; excitation rate 
$C^{e}_{jn}$), and deexcitations from levels $n$ to level $j$ ($n>j$; collisional deexcitation rates $C^{d}_{nj}$ and spontaneous decay 
rates $A_{nj}$) and from level $j$ to levels $m$ ($j>m$; collisional deexcitation and spontaneous decay rates $C^{d}_{jm}$ and 
$A_{jm}$, respectively). The ground level is populated/depopulated by the deexcitations/excitations from/to levels above it. Thus, the 
population of a level $j$ is given by the solution of \citep[][]{phillips2008}
\begin{eqnarray}
\label{excited_state}
\sum_{m<j}C_{mj}^{e}n_{_{e}}n_{Z,z,m}+\sum_{n>j}\left(A_{nj}+C_{nj}^{d}n_{e}\right)n_{Z,z,n} &-& \nonumber \\
n_{Z,z,j}\left[\sum_{j<n}C_{jn}^{e}n_{e}+\sum_{j>m}\left(A_{jm}+C_{jm}^{d}n_{e}\right)\right] & = & 0,
\end{eqnarray}
coupled to the equation of mass conservation
\begin{equation}
\sum_{j}n_{Z,z,j}=n_{Z,z}.
\end{equation}
In these equations $n_{Z,z,m}$, $n_{Z,z,j}$, $n_{Z,z,n}$ denote the population densities (in units of $cm^{-3}$) of levels $m$, $j$, and 
$n$, respectively; $n_{e}$ and $n_{Z,z}$ are the electron and ion densities. The units of the excitation/deexcitation rates are in 
cm$^{3}$ s$^{-1}$, while those of the spontaneous decay rates are in $s^{-1}$. 

The rates of collisional excitation from level $i$ to level $j$ ($i<j$) and deexcitation from level $j$ to level $i$ are determined by the 
average of the corresponding cross sections ($\sigma_{ij}(E_{e})$ and $\sigma_{ji}(E^{\prime}_{e})$ ) over all possible velocities of the 
incident electrons, that is, $C_{ij}^{e}=\langle \sigma_{ij}(E_{e}) v\rangle$ and $C_{ji}^{d}=\langle \sigma_{ji}(E^{\prime}_{e}) 
v^{\prime}\rangle$, respectively. The excitation cross section can be written in terms of the dimensionless collision strength 
$\Omega_{ij}$ through the expressions
\begin{equation}
\label{sigma1}
\sigma_{ij}(E_{e})=\pi a_{0}^{2}\frac{E_{H}}{\omega_{i}E_{e}}\Omega_{ij}(E_{e})=\pi 
a_{0}^{2}\frac{E_{e}}{\omega_{i}E_{ij}}\frac{\Omega_{ij}(E_{e})}{U}
\end{equation}
where $a_{0}$ is the Bohr radius, $E_{H}$ is the ground state energy of the hydrogen atom, $E_{e}$ is the energy of the incident 
electron, $\omega_{i}$ is the statistical weight of levels $i$ and $j$, $E_{ij}$ is the excitation energy between levels $i$, and 
$U=E_{e}/E_{ij}$ is the reduced electron energy. The collision strength is symmetrical with respect to the direct and the inverse 
processes, that is, $\Omega_{ij}(E_{e})=\Omega_{ji}(E^{\prime}_{e})$. Therefore, the deexcitation cross section can be written as
\begin{equation}
\sigma_{ji}(E^{\prime}_{e})=\pi a_{0}^{2}\frac{E^{\prime}_{e}}{\omega_{j}E_{ij}}\frac{\Omega_{ij}(E_{e})}{U^{\prime}}.
\end{equation}
In this expression $\omega_{j}$ is the statistical weight of level $j$, $E^{\prime}_{e}$ is the energy of the electron impacting on the 
electrons in level $j$ and and $U^{\prime}=E^{\prime}_{e}/E_{ij}$ is the reduced electron energy.

For electrons with a MB, $\kappa$ and $n$ distributions the rates of collisional excitation between levels $i$ and $j$ and deexcitation 
between levels $j$ and $i$ are given by\citep[see, for example,][]{dudik2014}
\begin{equation}
C^{e}_{ij}=8.629\times 10^{-6}\,T^{-1/2}\,\omega^{-1}_{i}\, e^{-y}\, \Upsilon_{ij}(T)
\end{equation}
and
\begin{equation}
C^{e}_{ji}=8.629\times 10^{-6}\,T^{-1/2}\,\omega^{-1}_{j}\, \rotatebox[origin=c]{180}{$\Upsilon$}_{ji}(T),
\end{equation}
respectively. In these expressions $y=E_{ij}/k_{B}T$, $\omega_{i}$ and $\omega_{j}$ are the statistical weights of levels $i$ and $j$, 
respectively, and $T$ is the temperature (in degrees K). The $\Upsilon_{ij}(T)$ (also known as the effective collision strength or Upsilon) and $\rotatebox[origin=c]{180}{$\Upsilon$}_{ji}(T)$ (also known as Downsilon) have different forms according to the electron 
distribution. They are given by
\begin{eqnarray}
\label{x1}
\Upsilon_{ij}(T) &= &y e^{y}\int_{1}^{+\infty}\Omega_{ij}(U)\, e^{-yU}dU\\
\rotatebox[origin=c]{180}{$\Upsilon$}_{ji}(T) &=& y \int_{0}^{+\infty}\Omega_{ji}(U^{\prime})\, e^{-yU^{\prime}}dU^{\prime}
\end{eqnarray}
for thermal,
\begin{eqnarray}
\label{x2}
\Upsilon_{ij}(T) &=& y e^{y}\int_{1}^{+\infty} \Omega_{ij}(U) \frac{A_{\kappa}}{\left(1+\frac{yU}{\kappa-3/2}\right)^{\kappa+1}}dU \\
\rotatebox[origin=c]{180}{$\Upsilon$}_{ji}(T)&=&y\int_{0}^{+\infty}\Omega_{ji}(U^{\prime})\, 
\frac{A_{\kappa}}{\left(1+\frac{yU^{\prime}}{\kappa-3/2}\right)^{\kappa+1}}dU^{\prime}
\end{eqnarray}
for $\kappa$, and
\begin{eqnarray}
\label{x3}
\Upsilon_{ij}(T) &= &y e^{y}B_{n}\int_{1}^{\infty} \Omega_{ij}(U) (yU)^{(n-1)/2}e^{-yU} dU\\
\rotatebox[origin=c]{180}{$\Upsilon$}_{ji}(T)&=&yB_{n}\int_{0}^{+\infty}\Omega_{ji}(U^{\prime})\,(yU^{\prime})^{(n-1)/2}e^{-yU^{\prime}} dU^{\prime}
\end{eqnarray}
for $n$ distributed electrons. It should be noticed from the above expressions that $\Upsilon_{ij}(T)\neq 
\rotatebox[origin=c]{180}{$\Upsilon$}_{ji}(T)$ for the non-Maxwellian distributions.

$\Omega_{ij}(U)$ can be written with the functional \citep{mewe1972,mewe1981,clarke1982,suno2006,dzifcakova2006,dzifcakova2015}
\begin{equation}
\label{x4}
\Omega_{ij}(U)=A_{0}+\sum_{n=1}^{n_{max}}\frac{A_{n}}{U^{n}}+D\ln U.
\end{equation}
Making use of the exponential integral of order $n$, $E_{n}(y)$ \citep{abramowitz1972}\footnote{
	The exponential integral of order $n$ is
	\begin{equation}
	\label{eqn8}
	E_{n}(y)=\int_{1}^{\infty}\frac{e^{-ty}}{t^{n}}dt, \mbox{ with }\, n>0.
	\end{equation}
	$E_{n+1}(y)$ can be determined through the recurrence formula
	\begin{equation}
	\label{eqn9}
	e^{y} E_{n+1}=\frac{1}{n}\left(1-y e^{y} E_{n}(y)\right).
	\end{equation}  
}
the integral in the RHS of (\ref{x1}) becomes
\begin{eqnarray}
\label{eqn10}
\int_{1}^{\infty} \Omega_{ij}(U)\, e^{-yU}dU &= &\int_{1}^{\infty} \left[A_{0}+\sum_{n=1}^{n_{max}}\frac{A_{n}}{U^{n}}+D\ln 
U\right]\,e^{-yU}dU \nonumber \\
&=& A_{0}\frac{e^{-y}}{y}+\sum_{n=1}^{n_{max}}\left[A_{n}E_{n}(y)\right]+\frac{D}{y}E_{1}(y)
\end{eqnarray}
and $\Upsilon_{ij}$ is given by
\begin{equation}
\label{eqn11}
\Upsilon_{ij} = A_{0}+y\sum_{n=1}^{n_{max}}\left(A_{n} \beta_{n}\right)+D\beta_{1},
\end{equation}
with $\beta_{n}=e^{y} E_{n}(y)$. We can use this result to fit the available $\Upsilon_{ij}$ and thus obtain the 
corresponding collision strength through the determination of the best fit parameters $A_{0}$, ..., $A_{n_{max}}$ and $D$ 
(\citet[][]{dzifcakova2006,dzifcakova2015}. This procedure is used because most of the theoretical collision strengths are not 
made available by their authors, instead what is available are the effective collision strengths calculated with the 
MB distribution. 

In the present calculations we assumed an atomic/ionic model composed of 30 levels and used the procedure discussed above with 
$n_{max}=8$ to determine the collision strengths of the transitions (see further details in de Avillez \& Anela 2018, in preparation). 

\subsection{Atomic data}
Electron impact ionization cross sections discussed in \citet{dere2007} and available in the Chianti database \citep[see, 
e.g.,][]{dere2009} are adopted.

The Maxwellian radiative recombination rate coefficients are taken from 
\citet{badnell2006ApJS}\footnote{http://amdpp.phys.strath.ac.uk/tamoc/DATA/} for all bare nuclei through Na-like ions recombining to 
H through Mg-like ions, \citet{altun2007} for Mg-like ions, \citet{abdel-naby2012} for Al-like ions, \citet{nikolic2010} for Ar-like 
ions, and \cite{badnell2006ApJ} for \fexiii-\fex ions. The Maxwellian dielectronic recombination rates are taken from 
\citet{badnell2006A&A} or H-like ions, \citet{bautista2007} for He-like ions, \citet{colgan2004,colgan2003} for Li and Be-like ions, 
\citet{altun2004,altun2006,altun2007} for B, Na and Mg-like ions, \citet{zat2003,zat2004a,zat2004b,zat2006} for C, O, F and Ne-like 
ions, \citet{mitnik2004} for N-like ions, \citet{abdel-naby2012} for Al-like ions, and \citet{nikolic2010} for Ar-like ions. We further used 
the data in the errata by \citet{altun2005} (for \nevi and \mgvii), \citet{zat2005b} (for \nitii), \citet{zat2005a} (for \neiii, \mgv, and \fexix).

Radiative and dielectronic recombination rates for \sulii, \suliii and \fevii are adopted from \citet{mazzotta1998A&AS..133..403M}, while 
for the remaining ions we adopt the radiative and dielectronic recombination rates derived with the unified electron-ion recombination 
method \citep{nahar1994PhRvA..49.1816N} and available at 
NORAD-Atomic-Data\footnote{http://www.astronomy.ohio-state.edu/$\sim$nahar}. 

The wavelengths, Einstein coefficients of spontaneous transitions and the effective collision strengths of transitions calculated for a MB 
electron distributions are taken from version 8.0.7 of the CHIANTI atomic 
database\footnote{http://www.chiantidatabase.org} \citep{delzanna2015}. 
\begin{figure}[thbp]
	\centering
	\includegraphics[width=0.85\hsize,angle=0]{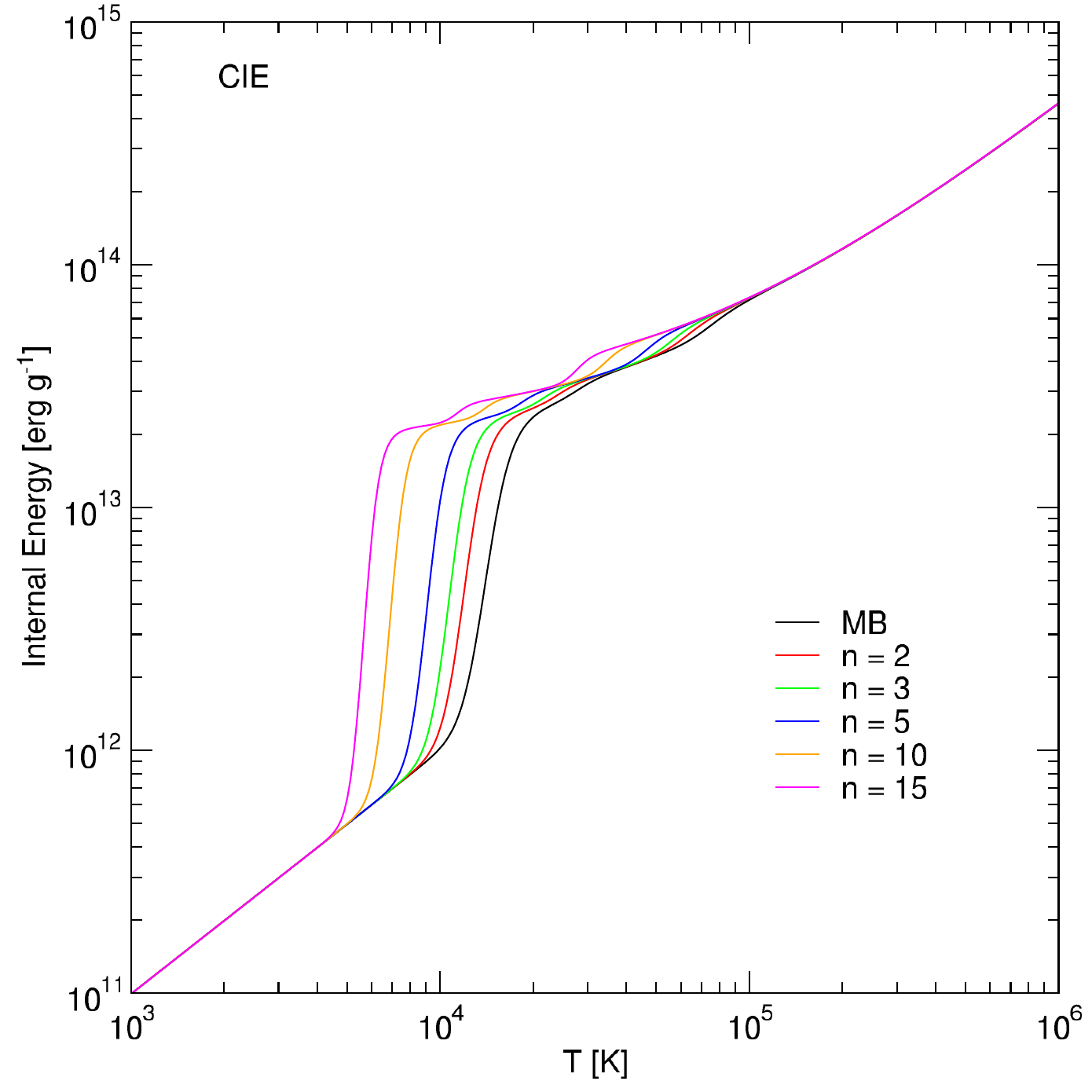}
	\includegraphics[width=0.85\hsize,angle=0]{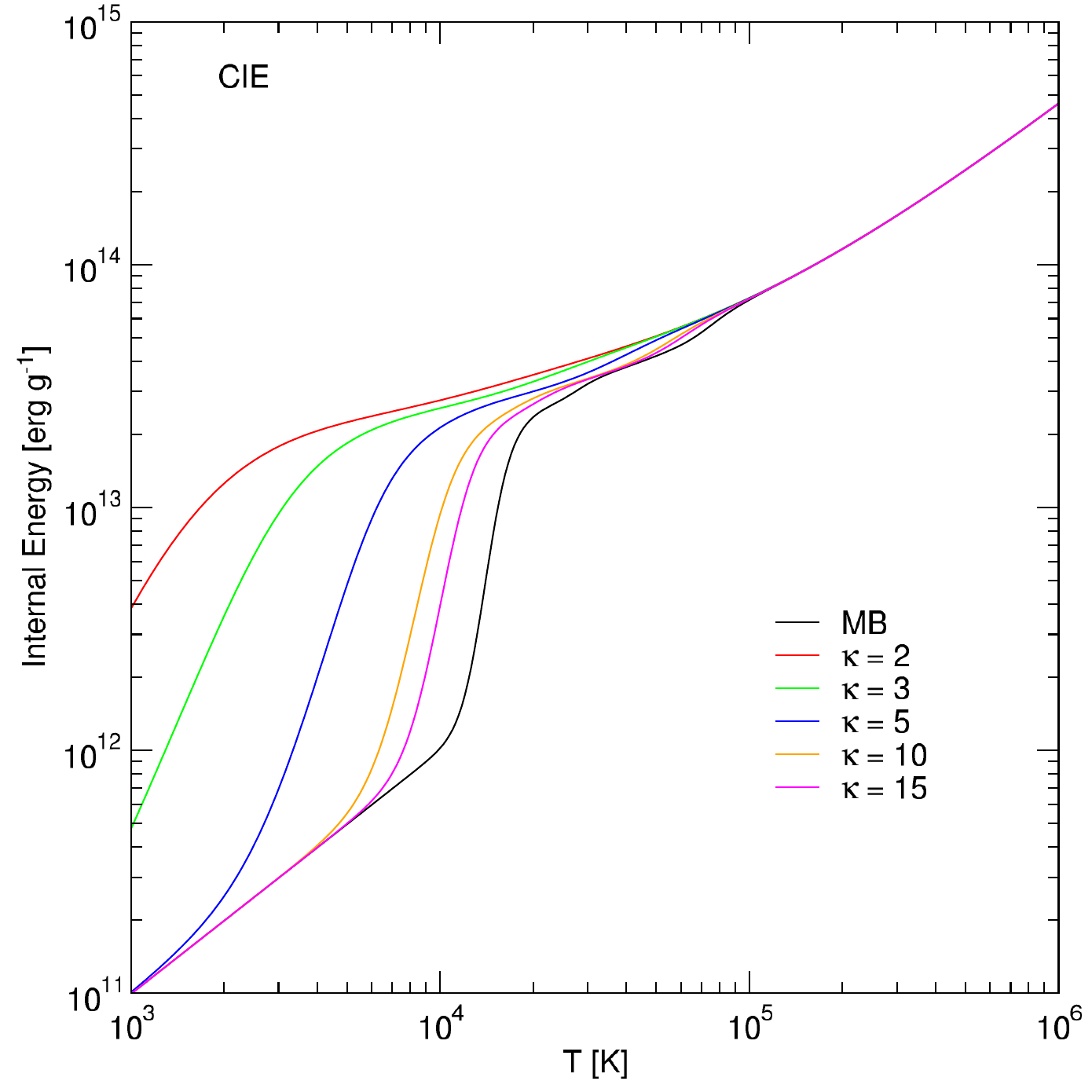}
	\caption{Total specific internal energy of the plasma for Maxwell-Boltzmann (black line in both panels), $n$ (top 
		panel) and $\kappa$ (bottom panel) distributions calculated for $n,\,\kappa=2,...,\, 15$.}
	\label{cie_internal_energy}
\end{figure}
\subsection{Numerical methods and calculations}
The ionic structure of the plasma is determined by solving the system of equations (\ref{ionfracseq}) at each temperature using a 
Gauss elimination method with scaled partial pivoting \citep{cheney2008} and a tolerance of $10^{-15}$. The numerical integrations 
in (\ref{ionization_rate}) are carried out with a precision of $10^{-15}$ using the double-exponential over a semi-finite interval 
method of \citet{takahasi1974} and \citet{mori2001}. We use a modified version (parallelized version) of the Numerical Automatic 
Integrator for Improper Integral package developed by T. Ooura\footnote{http://www.kurims.kyoto-u.ac.jp/\textasciitilde 
ooura/intde.html}.
\begin{figure}[thbp]
	\centering
	\includegraphics[width=0.85\hsize,angle=0]{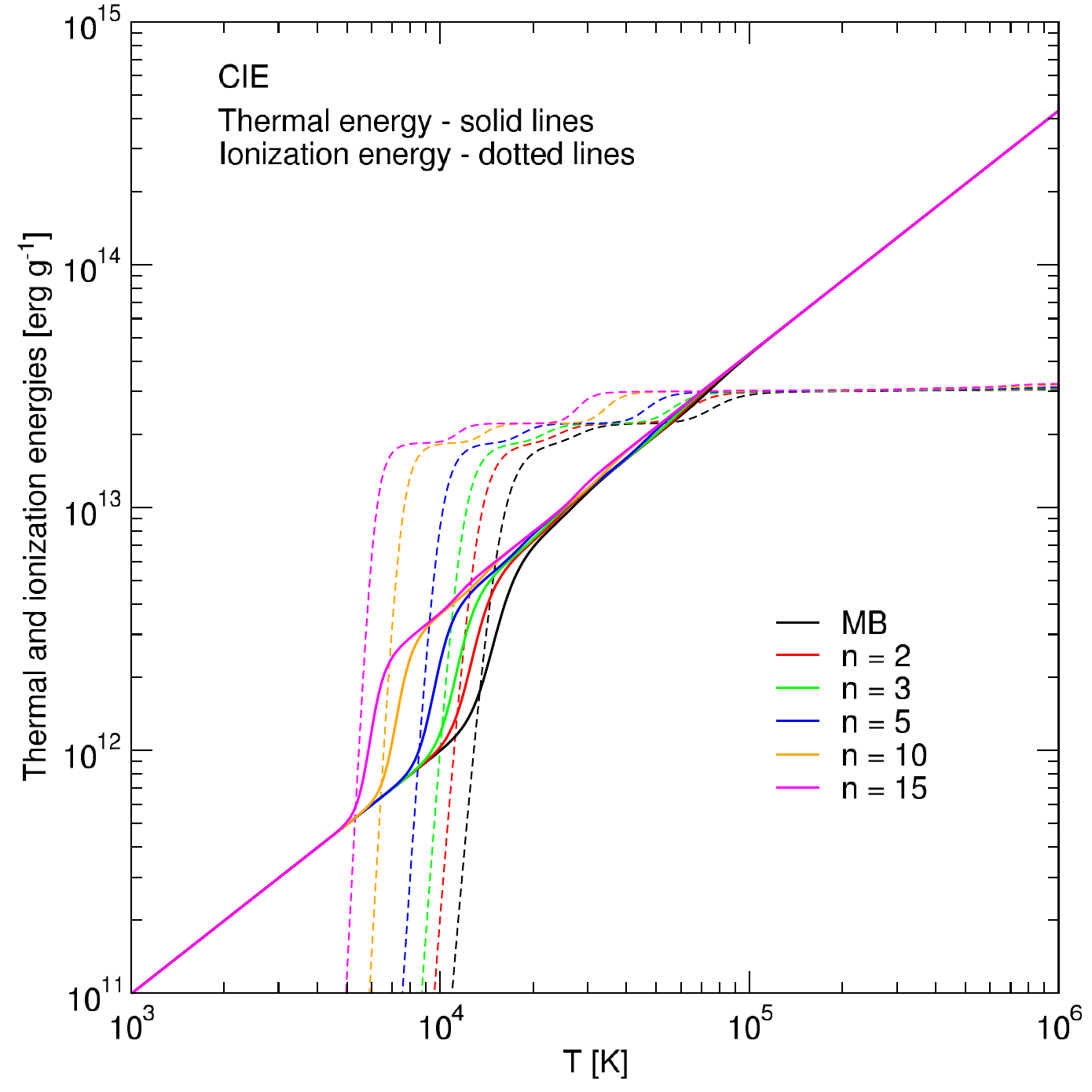}
	\includegraphics[width=0.85\hsize,angle=0]{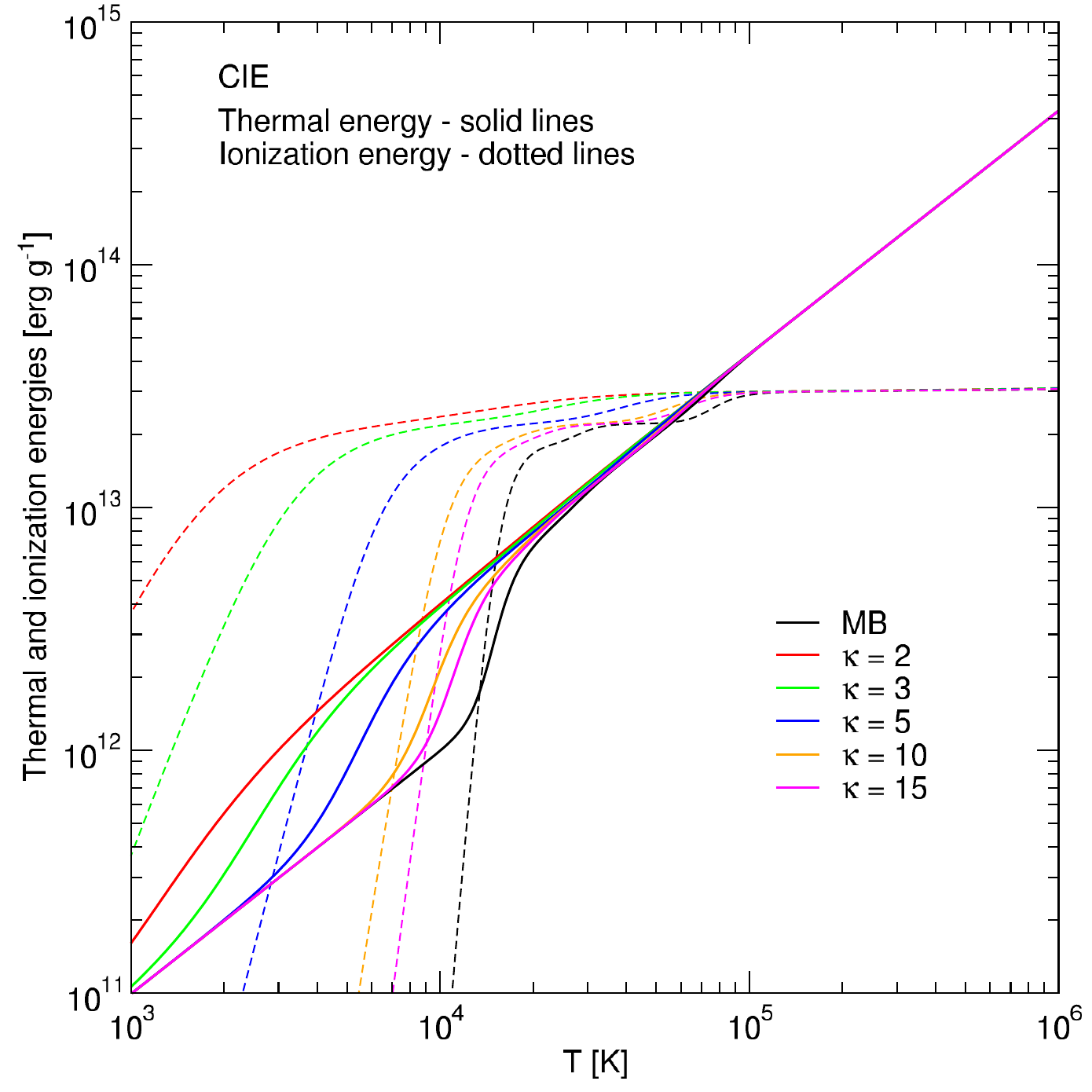}
	\caption{Specific thermal and ionization energies of the plasma for Maxwell-Boltzmann (black solid and dashed lines in both 
	panels), $n$ (top row) and $\kappa$ (bottom row) distributions with $n,\, \kappa=2,...,\,15$.}
	\label{cie_energies_separations}
\end{figure}
The calculations proceed as follows: follow the evolution of a gas parcel freely evolving isochorically under CIE from the initial 
conditions (fully ionized plasma at $10^{9}$ K with a density $n_{H}=1.0$ cm$^{-3}$); At each temperature the electron density, 
ionization and recombination rates and the ionic fractions are determined first. This is followed by the determination of the 
collisional excitation rates and of the populations of the excited states given by (\ref{excited_state}). Finally, the mean molecular 
weight, the pressure, the internal energy, the specific heat at constant volume $c_{v}$ and the gamma parameter are calculated.

\section{Results}

\subsection{Internal energy}
The total specific internal energy (erg g$^{-1}$) of a plasma evolving under collisional ionization equilibrium for the MB, $n$ and 
$\kappa$ distributions, using $n,\,\kappa=2$, 3, 5, 10 and 15 are displayed in Fig.~\ref{cie_internal_energy}, while its thermal and 
ionization components are shown in Fig.~\ref{cie_energies_separations}. The excitation component of the internal energy has less than a 
percent contribution to the total energy and thus, is not shown in the figures. The top and bottom panels in both figures refer to plasmas with, respectively, $n$ and $\kappa$ electron distributions. Maxwellian energies are shown by the black lines and dots in the top and bottom panels of both figures.

The total internal energy (as well as the thermal, ionization and excitation components) of the gas varies according to the parameters 
$\kappa$ and $n$. As $\kappa$ increases ($n$ increases) the internal energy approaches (moves away of) that obtained for a 
Maxwellian distribution of electrons. The largest variations in the total internal energy occur in $\kappa$ distributed plasmas and are 
mainly driven by the large variations (more than an order of magnitude with regard the thermal component for $T<10^{4}$ K)  in the 
ionization energy. In $n$ distributed plasmas the energies (total, thermal and ionization) profiles are similar  but with a shift to the 
left as $n$ increases (top panels of both figures). In the $n$-distributed plasma the dominance of the ionization energy over the 
thermal component only occurs down to the temperature at which the ionization energy has a sharp decrease. Below that 
temperature the thermal energy dominates (top panel of Fig.~\ref{cie_energies_separations}).
\begin{figure}[thbp]
	\centering
	\includegraphics[width=0.85\hsize,angle=0]{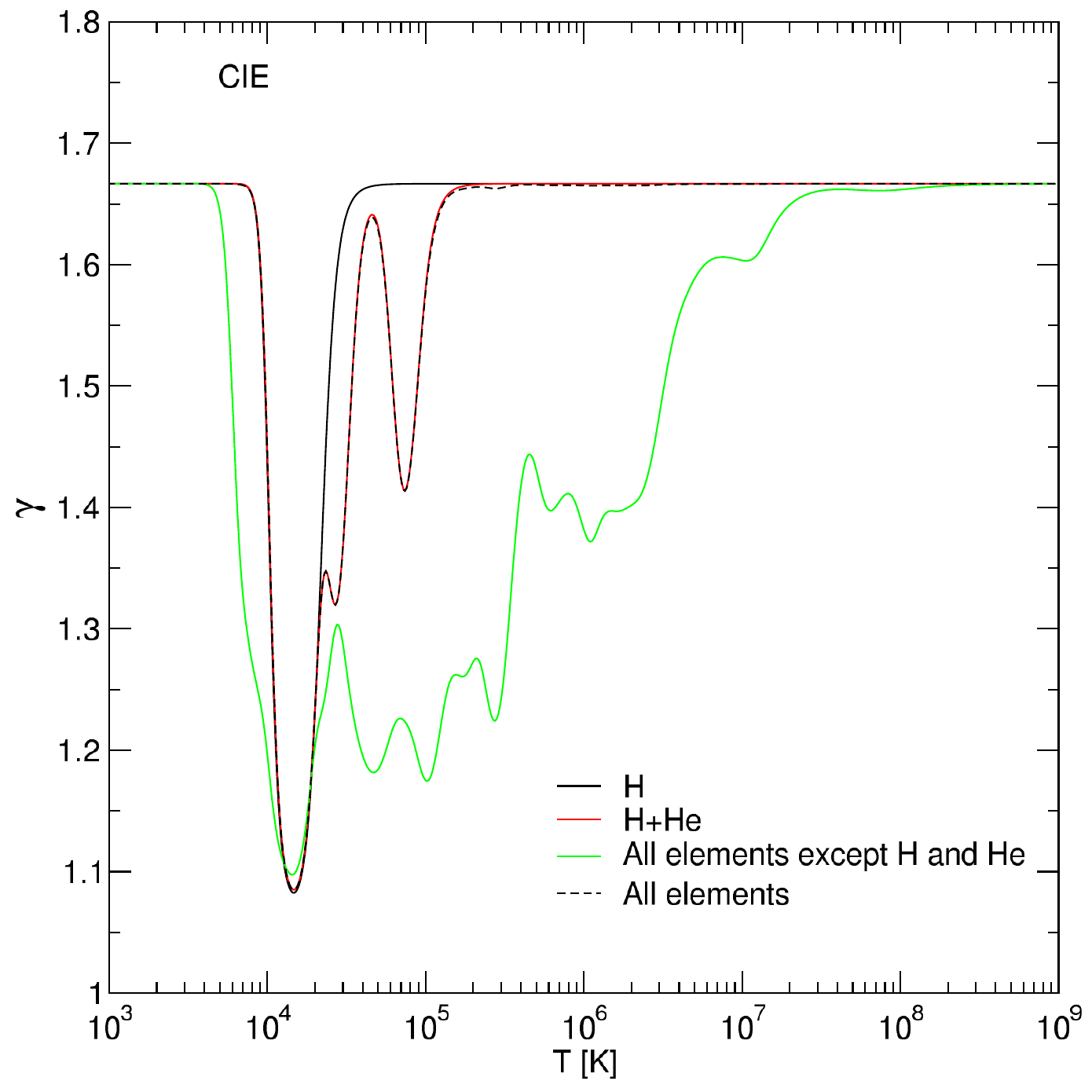}
	\caption{Evolution of the $\gamma$ parameter in different plasmas composed only of H (black line), H and He (red line), metals 
		(C, N, O, Ne, Mg, Si, S, and Fe; green line), and all elements (black dashed line). Note that H and He determine the evolution of 
		$\gamma$ for the calculations involving all the ten elements because the abundances of the metals are much smaller than 
		those of H and He.}
	\label{gamma_cie_mb}
\end{figure}

\subsection{The adiabatic parameter}

Fig.~\ref{gamma_cie_mb} displays the $\gamma$ parameter calculated for a monoatomic Maxwellian plasma composed of (1) 
H (black line), (2) H and He (red line), (3) metals (C, N, O, Ne, Mg, Si, S and Fe; green line), and (4) all elements (H, He and metals; 
black dashed line). The adiabatic parameter is 5/3 at $T\leq 8000$ K (cases 1,  2 and 4), at $T\le 3090$ K (case 3) and at $T\ge 
5\times 10^{4}$ K (case 1), $2\times 10^{5}$ K (case 2), $5\times 10^{6}$ K (case 4), and $4\times 10^{8}$ K (case 3). In the 
remaining temperatures $\gamma$ has substantial variations having values as low as 1.08 (cases 1, 2 and 4) and 1.1 (case 3). In 
the mixtures comprising H, He and metals the former two determine the value of $\gamma$ as they have abundances much larger 
than the remaining elements. Thus, the three transitions observed in case (4) are determined from left to right (from low to high 
temperatures) by the \hi, \hei and \heii ionizations, respectively. In case (3) the observed transitions in the $\gamma$ profile are due 
to the ionization of the most abundant elements among the metals.

\begin{figure}[thbp]
	\centering
	\includegraphics[width=0.85\hsize,angle=0]{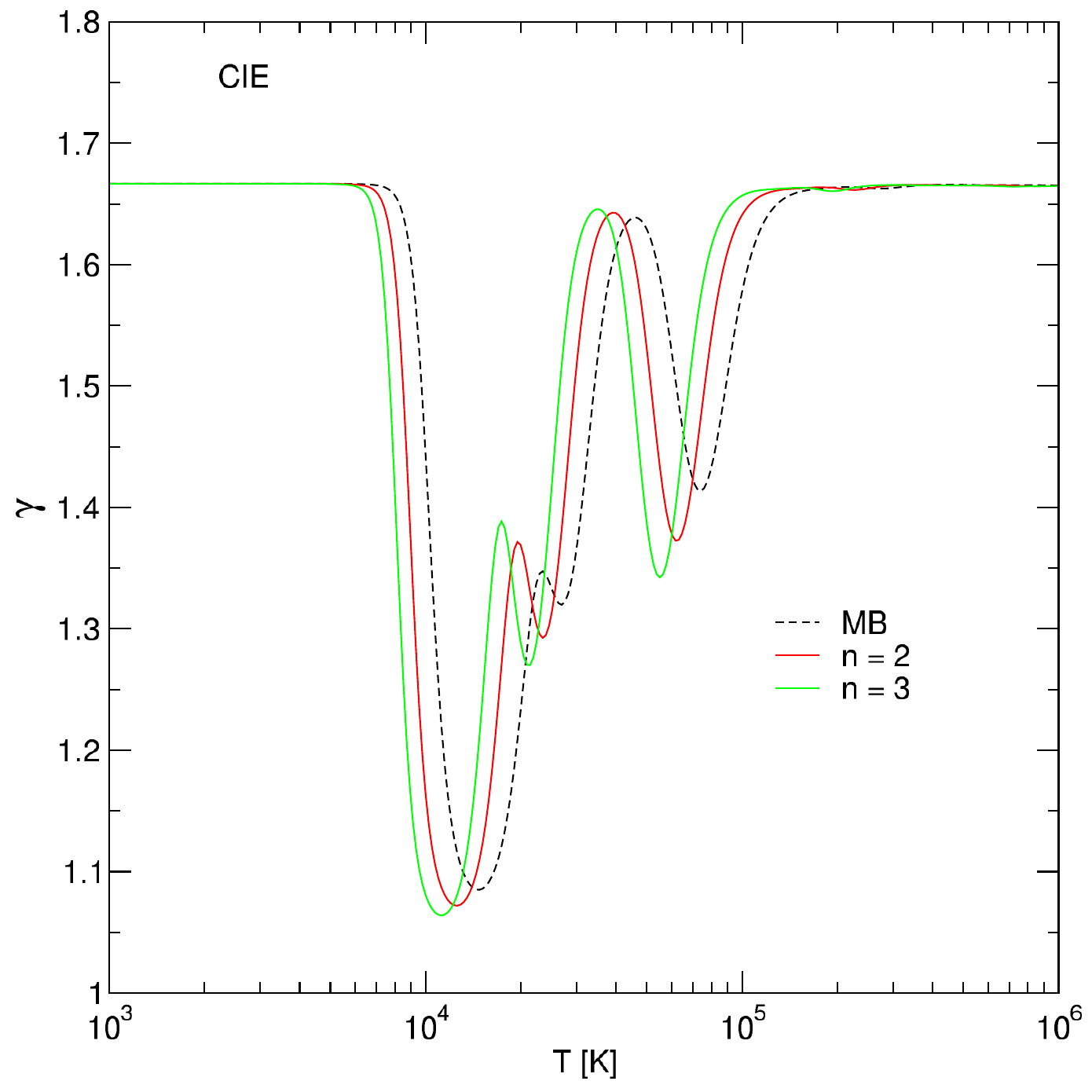}
	\includegraphics[width=0.85\hsize,angle=0]{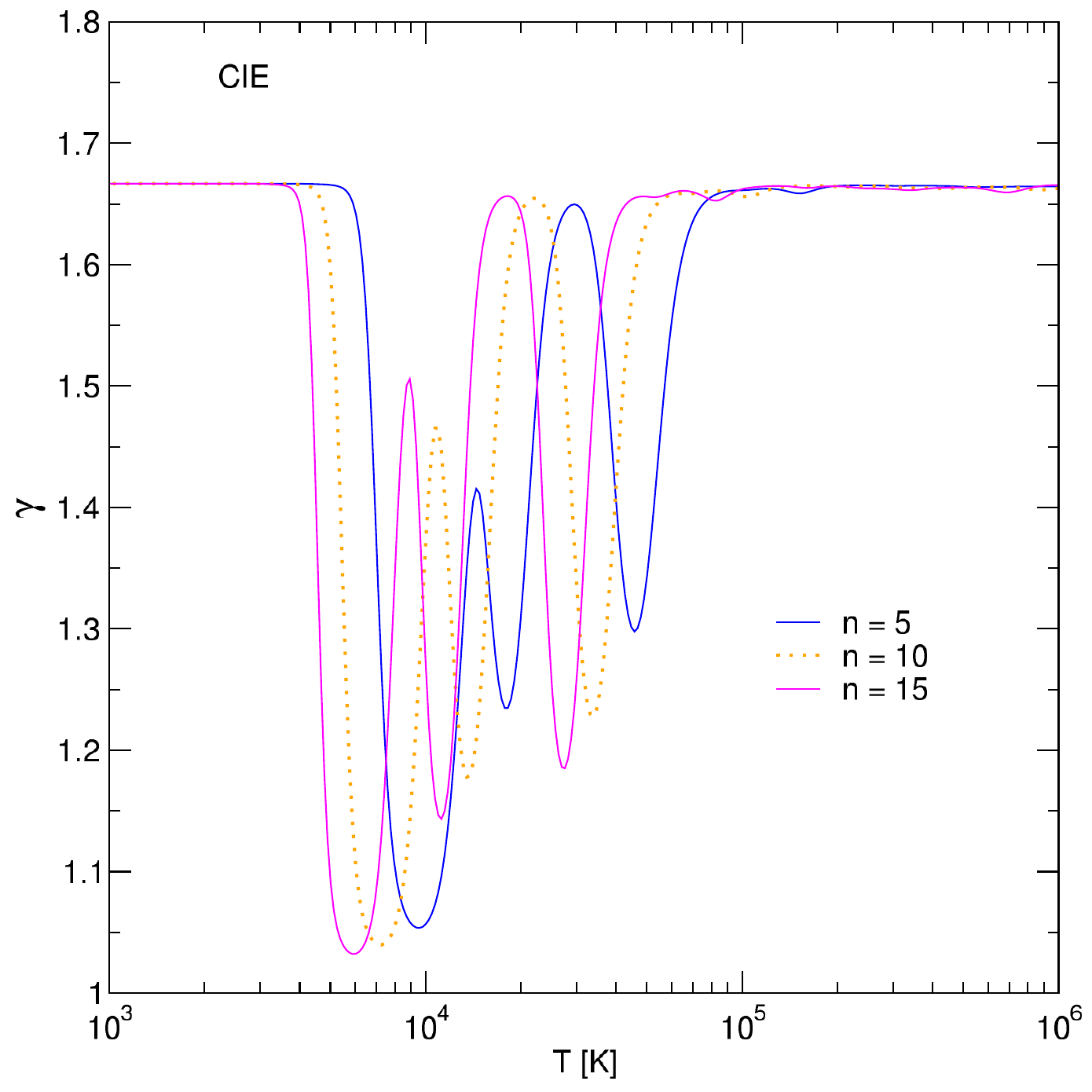}
	\includegraphics[width=0.85\hsize,angle=0]{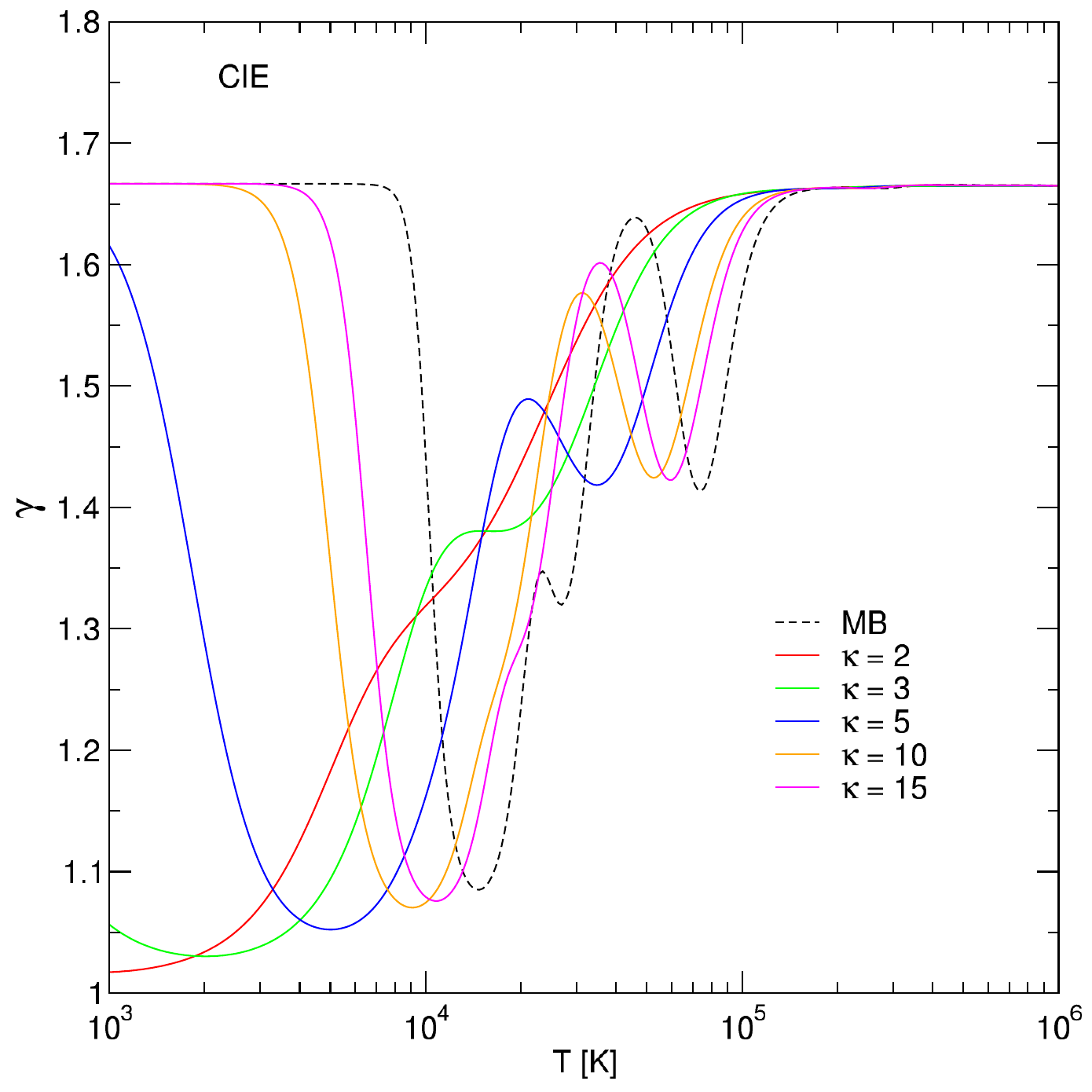}
	\caption{Evolution with temperature of the adiabatic parameter ($\gamma$) of a plasma evolving under CIE for 
		Maxwell-Boltzmann (black line in the top and bottom panels), $n$ (top and middle panels) and $\kappa$ (bottom panel) 
		distributions calculated for $n,\kappa=2$, 3, 5, 10 and 15.}
	\label{cie_gamma}
\end{figure}

Fig.~\ref{cie_gamma} displays the temperature variation of the adiabatic parameter in plasmas with $n$ (top and middle panels) and 
$\kappa$ (bottom panel) electron distributions and composed of a mixture of H, He and metals (case 4) for $n,\, \kappa=2,...,\,15$. 
The Maxwellian evolution of $\gamma$ ia shown by the black dashed lines. As $n$ increases $\gamma$ moves towards the left of 
the Maxwellian value with the same number of transitions (a result of the dominance of the ionization by H and He due to their 
abundances) but shifted to lower temperatures than those observed for lower $n$. The local minimum seen after each transition 
becomes deeper with increasing $n$. Therefore, the minimum value of $\gamma$ decreases from 1.09 for the Maxwellian 
distributed plasma to 1.025 for $n=15$ passing by 1.05 for $n=5$. That is, with increasing $n$ the plasma behaves almost 
isothermally at that range for temperatures.

\begin{figure}[thbp]
	\centering
	\includegraphics[width=0.85\hsize,angle=0]{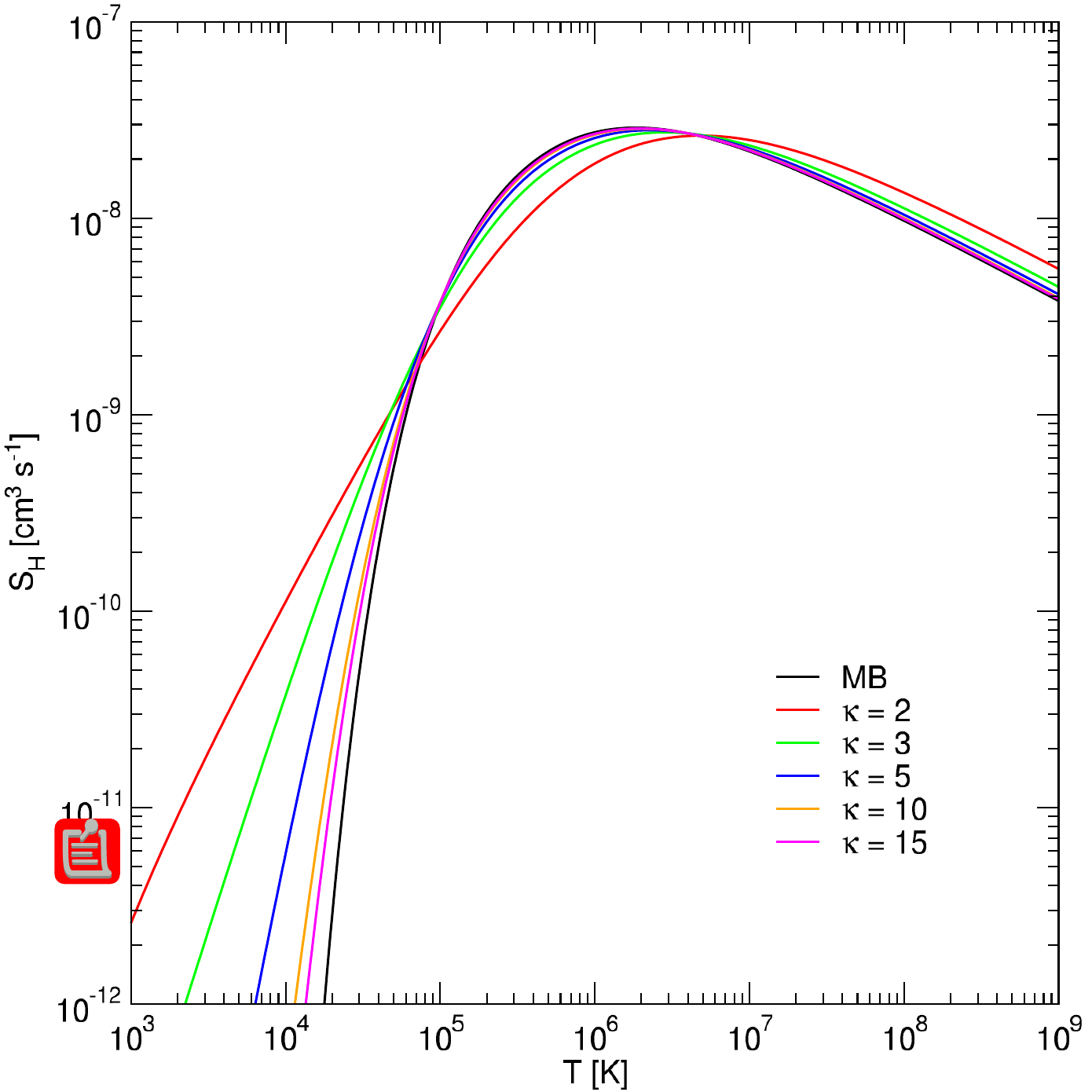}
	\includegraphics[width=0.85\hsize,angle=0]{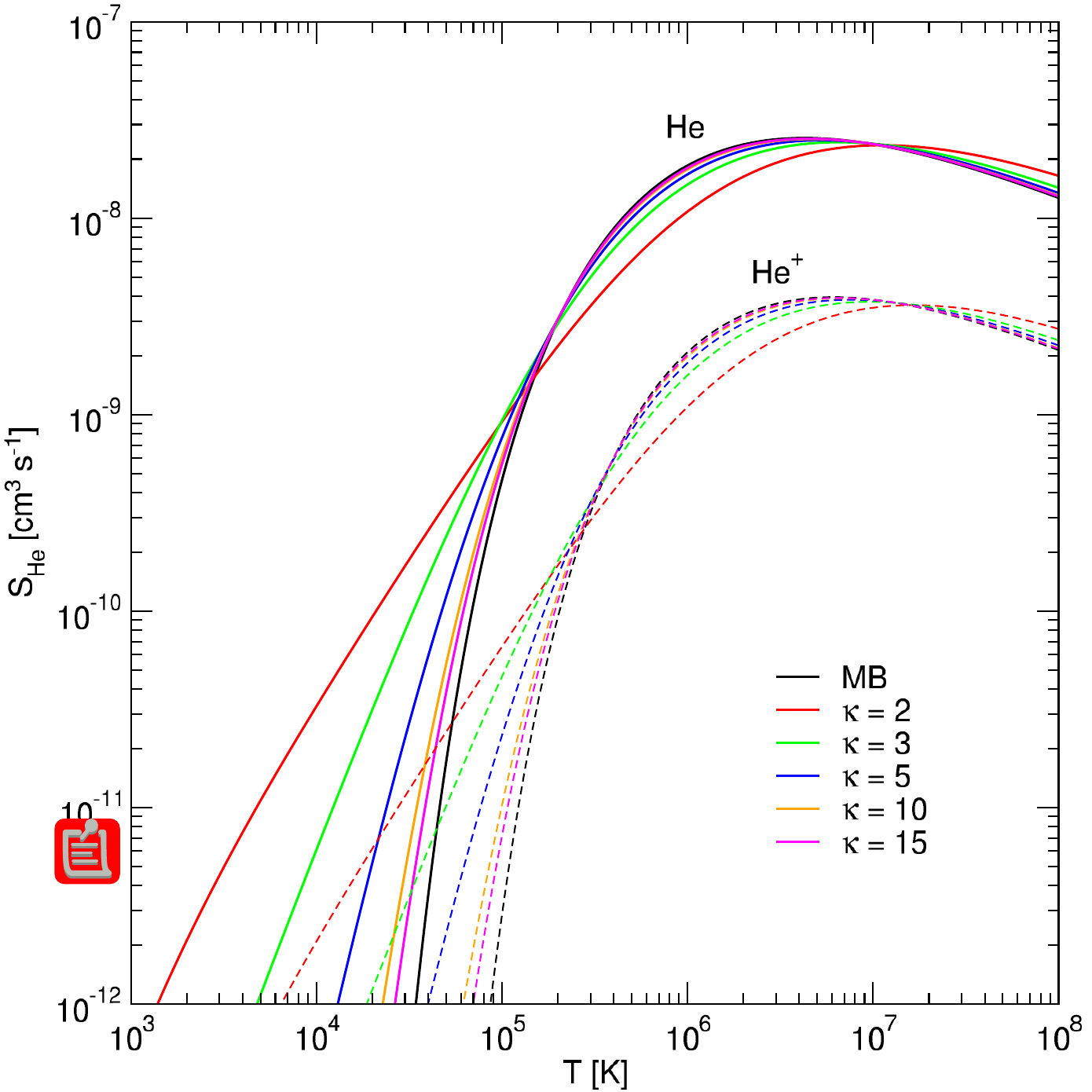}
	\caption{Variation with $\kappa$ of the ionization rates of H, He and He$^{+}$.}
	\label{eii_rates}
\end{figure}

In a $\kappa$ distributed plasma as $\kappa$ increases $\gamma$ approaches the Maxwellian value (bottom panel of 
Fig.~\ref{cie_gamma}). While the Maxwellian $\gamma$ has three transitions, the $\kappa$ distributed $\gamma$ only has three 
transitions for $\kappa>15$. Over the remaining values of $\kappa$ two transitions are seen for $\kappa>4$ and no transitions are 
seen at lower $\kappa$. The latter is a consequence of the early ionization of H and He when $\kappa$ has small values as can be 
seen in Fig.~\ref{eii_rates}, which displays the ionization rates of H, He and He$^{+}$, thereby affecting the internal energy of the 
plasma. As $\kappa$ increases the minimum value of $\gamma$ increases from almost 1.01 ($\kappa=2$) to 1.09 as $\kappa\to 
\infty$, that is, the Maxwellian value. $\gamma=5/3$ at $T>2\times 10^{5}$ K for all possible values of $\kappa$, while at 
$T<8000$ K $\gamma<5/3$ for $\kappa<7$. 

\section{Tables}

In the supplementary material tables (1 and 2) referring to the adiabatic parameter of the gas versus temperature (K) for the Maxwellian, 
$n$ and $\kappa$ distributions, with $n,\, \kappa=2, 3, 5, 10$ and 15, are provided. We note that $n=1$ corresponds to the Maxwellian 
distribution. More data can be provided by the corresponding author upon request.

\section{Final remarks}

The calculations presented in this paper comprise the determination of the $\gamma$ parameter in a plasma characterized by MB, 
$n$ and $\kappa$ electron distributions and evolving under collisional ionization equilibrium and cooling from $10^9$ K. Two 
particle processes (electron impact ionization including excitation-autoionization, radiative and dielectronic recombination and 
excitation) are considered in all the calculations. The populations of excited states are calculated assuming the coronal 
approximation. It is shown that the $\gamma$ parameter in an optically thin monoatomic plasma depends on its ionic state and 
most importantly on the contribution of the ionization energy to the total internal energy. The excitation energy has a negligible 
contribution to the total internal energy. Lookup tables of the temperature variation of $\gamma$ are provided as supplementary 
material and can be used in simulations  of solar, interstellar or intergalactic plasmas.

Although, CIE cooling functions are widely used in interstellar medium simulations caution should be exercised as CIE is only valid 
providing the cooling timescale ($\tau_{cool}$) of the plasma is larger than the recombination times scales of the different ions 
($\tau^{Z,z}_{rec}$), something that typically occurs at $T>10^{6}$ K. For lower temperatures, $\tau_{cool} <\tau_{rec}^{Z,z}$, and 
therefore, cooling and recombination are not synchronized, and the plasma appears overionized, if it is cooling \citep[for detailed 
discussions see, e.g.,][and references therein]{kafatos1973,shapiro1976,sutherland1993,schmutzler1993,gnat2007,avillez2010}. 
Consequently, the history of each gas parcel, evolving from different initial conditions, will have different ionic/radiative histories. 
Something that does not happen with CIE. 

Futhermore, \citet{avillez2012a} have shown in a muti-fluid time-dependent study of the dynamical and thermal evolutions 
of the interstellar gas\footnote{Hydrodynamical multi-fluid simulations of the interstellar medium in the disk and halo of the Milky  
Way including the time-dependent evolution of all atoms and ions associated to the ten most abundant elements in Nature (H, He, C, N, 
O, Ne, Mg, Si, S, and Fe).} that this history reflects the feedback between heating, cooling and the dynamics of the system. Therefore, 
even with the same initial conditions the gas parcels will have a different ionic/radiative history with the corresponding emissivities 
having an order of magnitude differences. 

We have shown that widely performed hydro- and magnetohydrodynamical simulations of optically thin plasmas like in the 
interstellar and intergalactic medium not only should take into account the time-dependent ionization structure 
\citep[see][]{avillez2012a}, but also the detailed distribution between internal and potential energies leading to different values of 
the ratio of specific heats, $\gamma$. In addition it should be checked during simulations, if MB equilibrium is established or 
if more generalized forms of electron distributions like $\kappa$ or $n$  are more applicable.

In a forthcoming paper we describe the evolution of the $\gamma$ parameter under non-equilibrium ionization conditions and with 
molecular contributions to the ionization state of the plasma as they affect the $\gamma$ parameter at low temperatures \citep[see, 
e.g.,][]{decampli1978,wuchterl1991,bodenheimer2013}.

\begin{acknowledgements}
We thank the anonymous referee for the comments improving the paper. This research was supported by the project Enabling Green 
E-science for the SKA Research Infrastructure (ENGAGE SKA), reference POCI-01-0145-FEDER-022217, funded by COMPETE 2020 and 
FCT, Portugal. D.B. acknowledges support from the \emph{Deut\-scheFor\-schungs\-ge\-mein\-schaft}, DFG project ISM-SPP 1573. The 
calculations were carried out at the ISM - Xeon Phi cluster of the Computational Astrophysics Group, University of \'Evora, acquired 
under project "Hybrid computing using accelerators \& coprocessors-modelling nature with a novell approach" (PI: M.A.), InAlentejo 
program, CCDRA, Portugal. 
\end{acknowledgements}

\bibliography{bibliography-gamma} 

\end{document}